\documentclass[twocolumn,floats,showpacs,superscriptaddress,pre]{revtex4}
\usepackage{amsmath,amssymb,bm}
\usepackage{graphics}
\usepackage{epsfig}
\usepackage{graphicx}

\begin{document}

\title{Covariance, relativity, and the proper mass of the universe in the
no-boundary wave function}
\author{Natalia Gorobey}
\affiliation{Peter the Great Saint Petersburg Polytechnic University, Polytekhnicheskaya
29, 195251, St. Petersburg, Russia}
\author{Alexander Lukyanenko}
\email{alex.lukyan@mail.ru}
\affiliation{Peter the Great Saint Petersburg Polytechnic University, Polytekhnicheskaya
29, 195251, St. Petersburg, Russia}
\author{A. V. Goltsev}
\affiliation{Ioffe Physical- Technical Institute, Polytekhnicheskaya 26, 195251, St.
Petersburg, Russia}

\begin{abstract}
A discrete class of privileged reference frames in a closed universe with
identical equations of motion for physical degrees of freedom was found. A
representation of the quantum state of the universe in a privileged
reference frame was obtained as a Euclidean functional integral with
no-boundary conditions. The boundary condition at the Pole, in addition to
the smoothness conditions, is the infinite asymptotic behavior of the Hubble
parameter. This makes it possible to regularize the functional integral by
changing the sign of the expansion energy of the universe. The proposed
construction also allows for the addition of a non-zero proper mass of the
universe.
\end{abstract}

\maketitle

%\author{Natalia Gorobey$^{1}$, Alexander Lukyanenko$^{1,\ast }$, and A.V.
%Goltsev$^{2}$}
%\email{$^{\ast }$alex.lukyan@mail.ru}
%\affiliation{$^{1}$Peter the Great Saint Petersburg Polytechnic University,
%Polytekhnicheskaya 29, 195251, St. Petersburg, Russia\\
%$^{2}$Ioffe Physical-Technical Institute, Polytekhnicheskaya 26, 195251, St.
%Petersburg, Russia}

%\date{\today }

%\pacs{}

%\begin{multicols}{2}
%\narrowtext

%%%%%%%%%%%%%%%%%%%%%%%%%%%%%%%%%%%%%%%%%%%%%%%%%%%%%%%%%%%%%%%%%%%%%%%

%%\bigskip

\section{\textbf{INTRODUCTION}}

Let's begin our discussion with the relationship between covariance and
relativity in physical theories. This issue was actively debated during the
development of Einstein's theory of gravity. If in the special theory of
relativity these two aspects -- covariance with respect to Lorentz
transformations and the independence of the form of physical laws from the
choice of an inertial frame of reference -- were in fact not distinguished,
then in the general theory of relativity it was recognized that these are
two different principles \cite{Fock}. General covariance is an obvious
formal requirement for writing physical laws in arbitrary space-time
coordinates. And the generalization of the principle of relativity to
non-inertial reference systems is possible under the condition that curved
space-time has some symmetries, and the reference systems, together with the
corresponding coordinate systems, realize this symmetry group. In the case
of an island distribution of gravitating masses with an asymptotically flat
spacetime at spatial infinity, one can speak of Lorentz covariance of the
theory if its predictions are also verified by a distant observer.
Observations within the island distribution are complicated by the effect of
the gravitational field on the measuring instruments. However, Fock
postulated a privileged role for harmonic coordinates, which serve as an
extension of Lorentz coordinates at spatial infinity into the region of a
strong gravitational field \cite{Fock}. Another variant of extrapolation is
considered in \cite{Fadd}. It is based on an elegant proof of the theorem of
positivity of the energy of the island mass distribution, proposed by Witten
\cite{Witt}. In this variant, the matching of measurements within the island
distribution with infinity is achieved by a special spinor field, which is a
solution of the 3D Dirac equation with appropriate boundary conditions at
infinity. In the case of a closed universe, the analogue of the theorem of
the positivity of the energy of the gravitational field is the
representation of the Hamiltonian function of the theory of gravity as the
difference between the energies of the physical degrees of freedom
(including gravitational waves) and the energy of expansion of the universe,
which is determined by the square of the above-mentioned 3D Dirac operator
\cite{Luk}. In development of this representation of the Hamiltonian
function of the universe, based on the Witten identity \cite{Witt}, in \cite%
{GL} a $W$-operator representation of gravitational constraints is
introduced, on the basis of which the algebra of constraints of the theory
of gravity can be modified by introducing additional independent parameters
-- the spectrum of values of the proper mass of the universe \cite{GLG}.

In this paper, we obtain a representation of gravitational constraints based
on the spectrum of the Witten operator $\widehat{W}$. The eigenfunctions of $%
\widehat{W}$, by analogy with \cite{Fadd}, define a class of privileged
reference frames in which the corresponding eigenvalues are Hamiltonian
functions. Being invariants of 3D diffeomorphisms, these Hamiltonian
functions can be modified by adding parameters of the proper mass of the
universe without violating the constraint algebra. For each privileged
reference frame, a no-boundary representation of the wave function of the
universe is given as a Euclidean functional integral. A regularization of
the functional integral by changing the sign of the expansion energy of the
universe is proposed.

The next section introduces a class of privileged reference frames in a
closed universe, defined by the spectrum of a Hermitian differential
operator equivalent to a system of local gravitational constraints. In the
third section, the quantum state of the universe in these reference frames
is represented by a Euclidean functional integral. In the fourth section,
the functional integral is regularized.

\section{PRIVILEGED FRAMES OF REFERENCE IN A CLOSED UNIVERSE}

The presence of privileged reference systems is a consequence of the
operator representation of gravitational connections on the spatial section $%
\Sigma $. The operator representation is based on the Witten identity \cite%
{Witt} in real integral form obtained in \cite{GL}:

\begin{equation}
\left( \chi ,\widehat{W}\eta \right) =\int_{\Sigma }d^{3}x\left[ N\left(
\chi ,\eta \right) \widetilde{\mathit{H}}_{0}+N^{i}\left( \chi ,\eta \right)
\widetilde{\mathit{H}}_{i}\right] ,  \label{1}
\end{equation}%
where $\chi ,\eta $ are arbitrary Dirac bi-spinors. Here $\widetilde{\mathit{%
H}}_{0},\widetilde{\mathit{H}}_{i}$ are the gravitational constraints in the
real canonical representation of Arnowitt, Deser and Misner \cite{ADM}, and
the operator $\widehat{W}$ on the space of bi-spinors is defined as the
difference of two positive definite operators, see \cite{GLG},

\begin{equation}
\widehat{W}=\Delta -\mathit{D}^{2}.  \label{2}
\end{equation}%
Here

\begin{equation}
\mathit{D}=i\sqrt{2}\left(
\begin{array}{c}
n_{\left. {}\right. A^{\prime }}^{A}\overline{\sigma }_{\left. {}\right.
B^{\prime }}^{kA^{\prime }}\overline{\triangledown }_{k}\overline{\mu }%
^{B^{\prime }} \\
n_{\left. {}\right. A}^{A^{\prime }}\sigma _{\left. {}\right.
B}^{kA}\triangledown _{k}\lambda ^{B}%
\end{array}%
\right) ,\eta =\left(
\begin{array}{c}
\lambda ^{A} \\
\overline{\mu }^{A^{\prime }}%
\end{array}%
\right)  \label{3}
\end{equation}
is 3D Dirac operator on the spatial section $\Sigma $, and $\Delta $ is the
sum of the Beltrami-Laplace operator with metric coefficients and the
energy-momentum tensor of the matter fields. $n_{AA^{\prime }}$ is an
arbitrary unitary spin matrix, the simplest of which is

\begin{equation}
n_{AA^{\prime }}=\frac{1}{\sqrt{2}}\left(
\begin{array}{cc}
1 & 0 \\
0 & 1%
\end{array}%
\right)  \label{4}
\end{equation}%
will be used in this work, and $\sigma _{AB}^{k}$ are the spin coefficients
of the metric $\beta _{kl}$ (see [6]). Operator Eq.(\ref{2}) will be further
called the Witten operator. With the simplest choice of $n_{AA^{\prime }}$
in the representation Eq.(\ref{4}), the spin coefficients of the metric are
real: $\overline{\sigma }^{kA^{\prime }B^{\prime }}=\sigma ^{kAB}$, where
the trait denotes complex conjugation, $A,A^{\prime }=0,1$. The covariant
derivative of the spinor field

\begin{equation}
\triangledown _{k}\lambda ^{A}=\partial _{k}\lambda ^{A}+A_{k\left.
{}\right. B}^{\left. {}\right. A}\lambda ^{B}  \label{5}
\end{equation}%
here it is constructed using Ashtekar's complex connection (taking into
account the reality condition) \cite{Ash},

\begin{equation}
A_{k\left. {}\right. B}^{\left. {}\right. A}=\Gamma _{k\left. {}\right.
B}^{\left. {}\right. A}\left( \sigma \right) +\frac{i}{\sqrt{2}}M_{k\left.
{}\right. B}^{\left. {}\right. A},  \label{6}
\end{equation}%
in which $M_{k}^{\left. {}\right. AB}$ are the canonical coordinates of the
gravity theory in the Ashtekar representation, conjugate to the spin
densities $\widetilde{\sigma }_{\left. {}\right. AB}^{k}$, which play the
role of canonical momenta, and $\Gamma _{k\left. {}\right. B}^{\left.
{}\right. A}\left( \sigma \right) $ is a real spin connection, which
uniquely determines the Christoffel connection \cite{Fran}. In our case, the
components of $M_{k}^{\left. {}\right. AB}$ are also real. The indices $%
A,A^{\prime }$ are raised and lowered using the antisymmetric spin tensor $%
\varepsilon _{AB}=-\varepsilon _{BA},\varepsilon _{01}=1$. The Hermitian
scalar product of bi-spinors is defined as

\begin{equation}
\left( \chi ,\eta \right) =\int_{\Sigma }\sqrt{\beta }d^{3}xn_{AA^{\prime }}%
\left[ \mu _{1}^{A}\overline{\mu }_{2}^{A^{\prime }}+\overline{\lambda }%
_{1}^{A^{\prime }}\lambda _{2}^{A}\right] ,  \label{7}
\end{equation}%
where $\beta =\det \beta _{kl}$, and $\beta _{kl}$ is the metric on the 3D
spatial section $\Sigma $. The Witten operator is Hermitian with respect to
this scalar product. Finally, the coefficients on the right-hand side of Eq.(%
\ref{1}) (the lapse and shift functions \cite{MTW}) are equal to:

\begin{equation}
N\left( \chi ,\eta \right) =-\frac{1}{8}n_{AA^{\prime }}\left[ \mu _{1}^{A}%
\overline{\mu }_{2}^{A^{\prime }}+\overline{\lambda }_{1}^{A^{\prime
}}\lambda _{2}^{A}\right] ,  \label{8}
\end{equation}

\begin{equation}
N^{k}\left( \chi ,\eta \right) =\frac{i}{4}\sigma _{\left. {}\right. AB}^{k}%
\left[ \mu _{1}^{A}\mu _{2}^{+B}+\lambda _{1}^{+B}\lambda _{2}^{A}\right] ,
\label{9}
\end{equation}%
and

\begin{equation}
\lambda ^{+A}=\sqrt{2}n_{\left. {}\right. A^{\prime }}^{A}\overline{\lambda }%
^{A^{\prime }}.  \label{10}
\end{equation}

Let us consider the secular equation for the Witten operator:

\begin{equation}
\widehat{W}\chi =W\chi .  \label{11}
\end{equation}%
The eigenvalue can be represented as:

\begin{equation}
W=\frac{\left( \chi ,\widehat{W}\chi \right) }{\left\vert \left\vert \chi
\right\vert \right\vert ^{2}},  \label{12}
\end{equation}%
where $\chi $ is the corresponding eigenvector. These quantities are
functionals of all fundamental canonical variables of gravity theory.
Obviously, the eigenvalue $W$ is invariant under 3D coordinate
transformations on $\Sigma $, since the secular equation is covariant. This
means that $W$ PB (Poisson brackets)-commutes with the momentum constraints $%
\widetilde{\mathit{H}}_{i}$ -- the generators of 3D diffeomorphisms,

\begin{equation}
\left\{ W,\widetilde{\mathit{H}}_{i}\left( x\right) \right\} =0,  \label{13}
\end{equation}%
while the generators of diffeomorphisms themselves form a closed algebra:

\begin{equation}
\left\{ \widetilde{\mathit{H}}_{i}\left( x\right) ,\widetilde{\mathit{H}}%
_{k}\left( y\right) \right\} =\varepsilon _{ikl}\widetilde{\mathit{H}}%
_{l}\left( x\right) \delta ^{3}\left( x-y\right) ,  \label{14}
\end{equation}%
where $\varepsilon _{ikl}$ is a completely antisymmetric 3D tensor. It
remains to find the PB -commutator of two eigenvalues of the Witten
operator. It is easy to find using identity Eq.(\ref{1}) and the commutation
relations of the ADM constraint algebra \cite{Fran}. Introducing the
normalized eigenvectors of the Witten operator $\widetilde{\chi }=\chi
/\left\vert \left\vert \chi \right\vert \right\vert $, we write the result:

\begin{equation}
\left\{ W_{1},W_{2}\right\} =\int_{\Sigma }d^{3}xC_{12}^{i}\left( x\right)
\widetilde{\mathit{H}}_{i}\left( x\right) ,  \label{15}
\end{equation}%
where

\begin{equation}
C_{12}^{i}=\left( \partial _{k}N_{1}N_{2}-N_{1}\partial _{k}N_{2}\right)
\beta ^{ik},  \label{16}
\end{equation}%
while $N_{1,2}=N\left( \chi _{1,2},\chi _{1,2}\right) $, and

\begin{equation}
\partial _{k}N=N\left( \triangledown _{k}\chi ,\chi \right) +N\left( \chi
,\triangledown _{k}\chi \right) ,  \label{17}
\end{equation}%
where the covariant derivative is calculated with the Ashtekar connection
Eq.(\ref{6}). Thus, the quantities $W,\widetilde{\mathit{H}}_{i}\left(
x\right) $ form a closed algebra with respect to the Poisson brackets. Note
that the eigenvalues $W$ are absent from the right-hand sides of all
commutation relations. This allows us to modify the theory by adding
arbitrary constants to the eigenvalues of the Witten operator, which form
the spectrum of the proper mass of the universe in different reference
frames. This ambiguity of the classical theory of gravity can be interpreted
differently. After quantization, the Witten operator becomes a double
operator $\widehat{\widehat{W}}$, which is defined on the elements $\chi
\otimes \Psi $ of the Cartesian product of two spaces - the space of Dirac
bi-spinors on $\Sigma $ and the Hilbert space of states of the universe.
Then the proper mass of the universe is formed by the eigenvalues of this
double operator, and the arbitrariness in its definition disappears. In this
form of QTG, the quantum state of the universe $\Psi $ is defined together
with the reference frame, which is given by the bi-spinor $\chi $.

Let us return, however, to the standard quantization procedure, in which the
original function is the classical Hamiltonian function. In the new
representation of the constraint algebra Eqs.(\ref{13}), (\ref{14}), (\ref%
{15}), the main object is (any) eigenvalue $W=\left( \widetilde{\chi },%
\widehat{W}\widetilde{\chi }\right) $ of the Witten operator, where $%
\widetilde{\chi }$ is the corresponding normalized eigenvector. Here, the
set of eigenvalues replaces the Hamiltonian constraints of the ADM defined
at each point of $\Sigma $. The latter are generators of the multi-fingered
time shift \cite{MTW} on the spatial section of $\Sigma $, and each
eigenvalue $W$ now serves as a generator of the proper time shift in a
privileged reference frame defined by the corresponding eigenvector $%
\widetilde{\chi }$.

Thus, we have made the transition from covariance to relativity in a closed
universe at the classical level. Covariance (the independence of the form of
the equations of motion from an arbitrary time transformation at each point $%
\Sigma $) is replaced by the independence of the form of physical laws from
the choice of a reference frame from the class of privileged ones determined
by the eigenvectors $\widetilde{\chi }$ of the Witten operator. By fixing a
reference frame (eigenvector $\widetilde{\chi }$), we define the Hamiltonian
function of the universe $W=\left( \widetilde{\chi },\widehat{W}\widetilde{%
\chi }\right) $ in this reference frame. The invariance of the theory under
arbitrary transformations of spatial coordinates in each reference frame now
does not require additional conditions in the form of constraints $%
\widetilde{\mathit{H}}_{i}\left( x\right) $. This is a consequence of the 3D
covariance of the secular equation Eq.(\ref{11}). Below, we show that the
spectrum of the Witten operator and, accordingly, the class of privileged
reference frames are discrete.

\section{NO-BOUNDARY WAVE FUNCTION OF THE UNIVERSE}

We fix a reference frame corresponding to the eigenvector $\widetilde{\chi }$
of the Witten operator and proceed to quantum theory in this reference
frame. We use the functional integral formalism for the invariant propagator
of the Schr\"{o}dinger equation \cite{FV,BV}. However, we do not need to
fully utilize the Batalin-Fradkin-Vilkovisky theorem for the invariant
propagator. In the simplest case of a relativistic particle, the result of
its application reduces to the following \cite{Gov}:

\begin{equation}
\mathit{k}=\int_{0}^{\infty }dT\int \prod\limits_{t=0}^{T}d^{4}xd^{4}p\exp %
\left[ -\frac{1}{\hslash }I_{E}\right] ,  \label{18}
\end{equation}%
where

\begin{equation}
I_{E}=\int_{0}^{T}dt\left[ \left( p\overset{\cdot }{x}\right) _{E}-\left(
p_{E}^{2}+m^{2}\right) \right] .  \label{19}
\end{equation}%
is the particle action in canonical Euclidean form. We immediately use
Euclidean quantum theory, as is necessary in the case of the no-boundary
wave function. Note that here

\begin{eqnarray}
p_{E}^{2} &=&p_{0}^{2}+p_{1}^{2}+p_{2}^{2}+p_{3}^{2},  \notag \\
\left( p\overset{\cdot }{x}\right) _{E} &=&p_{0}\overset{\cdot }{x}^{0}+p_{1}%
\overset{\cdot }{x}^{1}+p_{2}\overset{\cdot }{x}^{2}+p_{3}\overset{\cdot }{x}%
^{3}  \label{20}
\end{eqnarray}%
which is obtained according to the following rules of the Wick rotation:

\begin{equation}
t\rightarrow it,x^{0}\rightarrow ix^{0},p_{0}\rightarrow
p_{0},x^{k}\rightarrow x^{k},p_{k}\rightarrow ip_{k}.  \label{21}
\end{equation}

The theory of gravity in a fixed reference frame $\widetilde{\chi }$ differs
from a relativistic particle only in that the number of "spatial" degrees of
freedom of the universe is infinite. However, the proper time of the
universe in the reference frame $\widetilde{\chi }$ is one. Therefore, we
can use formula Eq.(\ref{18}) for the invariant propagator of the universe
and write:

\begin{equation}
\mathit{k}_{GR}=\int_{0}^{\infty }dT\int \prod\limits_{t=0}^{T}D\mathit{Q}D%
\mathit{P}\exp \left[ -\frac{1}{\hslash }I_{EGR}\right] ,  \label{22}
\end{equation}%
where

\begin{equation}
I_{EGR}=\int_{0}^{T}dt\left[ \left( \sum \mathit{P}\overset{\cdot }{\mathit{Q%
}}\right) _{E}-\left( \left( \widetilde{\chi },\widehat{W}\widetilde{\chi }%
\right) _{E}+m_{\chi }^{2}\right) \right]  \label{23}
\end{equation}%
is the action of the universe in Euclidean canonical form, where for
generality we have added the proper mass of the universe in a given
reference frame, and ($\mathit{Q}$\textit{,}$\mathit{P}$) are all canonical
variables of geometry and matter fields. The constraints corresponding to
the internal gauge symmetries of the matter fields should also be added to
the Hamiltonian function, but we have excluded 3D diffeomorphisms. By taking
a certain solution $\widetilde{\chi }$ of the covariant secular equation Eq.(%
\ref{11}), we fix the spatial coordinates on $\Sigma $. Note that $%
\widetilde{\chi }$ is canonically "neutral." This means, in particular, that
when eliminating the canonical momenta in Eq.(\ref{23}), $\widetilde{\chi }$
should not be differentiated with respect to $\mathit{P}$.

Now we define the boundary conditions for the action Eq.(\ref{23}) and the
propagator Eq.(\ref{22}), which will correspond to our understanding of the
no-boundary wave function of the universe. In the original definition of
Hartle-Hawking \cite{HH}, a compact 4D space is considered with a sequence
of "spatial" sections $\Sigma $, contracting to a single point -- the Pole.
Taking into account the $3+1$-splitting of the 4D geometry of the ADM, the
boundary 3D metric $\beta _{kl}$ and matter fields $\phi _{\alpha }$ are
specified at one extreme section, and at the Pole (at $t=0$) the smoothness
conditions are:

\begin{equation}
\beta =\det \beta _{kl}=0,\frac{d}{dt}\left( \beta _{kl}/\beta ^{1/3}\right)
=0,\frac{d}{dt}\phi _{\alpha }=0.  \label{24}
\end{equation}%
Thus, $\beta $ plays the role of the (multi-arrow) time parameter in this
cross-section congruence, and the Pole serves as the origin. The functional
integration in Eq.(\ref{22}) is carried out over all trajectories in the
configuration space between the given boundary values (we assume that the
canonical momenta are excluded first). However, such a parameterization of
the cross-section congruence turns out to be unsatisfactory from the point
of view of the convergence of integral Eq.(\ref{22}) \cite{Haw}. The entire
convergence problem lies in the canonical pair ($M$,$\sqrt{\beta }$) with
the commutator:

\begin{equation}
\left\{ M\left( x\right) ,\sqrt{\beta \left( y\right) }\right\} =\frac{3}{2}%
\delta ^{3}\left( x-y\right) .  \label{25}
\end{equation}%
In the complex canonical representation of Ashtekar's theory of gravity,
given the reality condition Eq.(\ref{6}), it is $M_{kB}^{A}\sigma _{A}^{kB}$
that plays the role of the canonical coordinate in this pair and should be
used as the congruence parameter. Note that in the case of a homogeneous
universe, in accordance with Eq.(\ref{25}), $M$ is proportional to the
Hubble parameter $\overset{\cdot }{a}/a$, where $a$ is the radius of the
universe.

\section{REGULARIZATION OF THE FUNCTIONAL INTEGRAL}

To regularize the functional integral Eq.(\ref{22}), we perform a Wick
rotation in the action Eq.(\ref{23}), analogous to Eq.(\ref{21}), by
choosing as the $x^{0}$ coordinate of the Minkowski space the set of
canonical coordinates $M(x)$ in the configuration space of the universe.
These quantities are contained in the Dirac operator Eq.(\ref{3}), but also
in the first part $\Delta $ of the Witten operator. The Dirac operator can
be completely `extracted' from $\Delta $ (see \cite{GLG}), so that in the
modified Witten operator

\begin{equation}
W^{\prime }=\Delta ^{\prime }-\frac{11}{9}\mathit{D}^{2},  \label{26}
\end{equation}%
$\Delta ^{\prime }$ does not contain the canonical coordinates of $M(x)$. We
will further assume that the Hamiltonian function is $\left( \widetilde{\chi
},\widehat{W}\widetilde{\chi }\right) _{E}$ with the modified Witten
operator, omitting the prime.

Let us consider the secular equation for the Dirac operator,

\begin{equation}
\mathit{D}\eta =\mathit{d}\eta .  \label{27}
\end{equation}%
The eigenvalue $\mathit{d}$ can be represented as:

\begin{eqnarray}
\mathit{d} &=&\left( \widetilde{\eta },\mathit{D}\widetilde{\eta }\right) =i%
\sqrt{2}\int_{\Sigma }d^{3}x\left[ \mu _{B}\widetilde{\sigma }_{\left.
{}\right. C}^{kB}\overset{\left( 3\right) }{\triangledown }_{k}\lambda
^{C}\right.  \notag \\
&&\left. +\frac{i}{\sqrt{2}}\sqrt{\beta }M\mu _{B}\lambda ^{B}+\overline{%
\lambda }_{B^{\prime }}\widetilde{\overline{\sigma }}_{\left. {}\right.
C^{\prime }}^{kB^{\prime }}\overset{\left( 3\right) }{\overline{%
\triangledown }}_{k}\overline{\mu }^{C^{\prime }}-\right.  \notag \\
&&\left. -\frac{i}{\sqrt{2}}\sqrt{\beta }M\overline{\lambda }_{B^{\prime }}%
\overline{\mu }^{B^{\prime }}\right] ,  \label{28}
\end{eqnarray}%
where $\widetilde{\eta }$ is a normalized eigenvector with components $%
\lambda ^{A},\overline{\mu }^{A^{\prime }}$, and

\begin{equation}
\triangledown _{k}\lambda ^{A}=\overset{\left( 3\right) }{\triangledown }%
_{k}\lambda ^{A}+\frac{i}{\sqrt{2}}M_{kB}^{A}\lambda ^{B},  \label{29}
\end{equation}%
where the Ashtekar connection Eq.(\ref{6}) is separated into real and
imaginary parts. We perform a Wick rotation here, $M\rightarrow iM$.
Simultaneously, we integrate by parts in the second term of the second row
in Eq.(\ref{28}), resulting in:

\begin{equation}
-\overset{\left( 3\right) }{\overline{\triangledown }}_{k}\overline{\lambda }%
_{B^{\prime }}\widetilde{\overline{\sigma }}_{\left. {}\right. C^{\prime
}}^{kB^{\prime }}\overline{\mu }^{C^{\prime }}=-\overset{\left( 3\right) }{%
\overline{\triangledown }}_{k}\overline{\lambda }^{B^{\prime }}\widetilde{%
\overline{\sigma }}_{\left. B^{\prime }\right. }^{k\left. C^{\prime }\right.
}\overline{\mu }_{C^{\prime }}.  \label{30}
\end{equation}%
Then, taking into account the realness of $\widetilde{\sigma }_{\left.
{}\right. C}^{kB},M_{kB}^{A}$, if we assume that the components $\lambda
^{A},\overline{\mu }^{A^{\prime }}$ of the eigenvector $\widetilde{\eta }$
are also real (the sign of complex conjugation is superfluous, but we leave
it), then the first and third terms under the integral sign in Eq.(\ref{28})
are cancelled out, and the remaining two give:

\begin{equation}
\mathit{d}=-2i\int_{\Sigma }d^{3}x\sqrt{\beta }\mu _{B}\lambda ^{B}M,
\label{31}
\end{equation}%
where it is taken into account that

\begin{equation}
\overline{\lambda }_{B^{\prime }}\overline{\mu }^{B^{\prime }}=\lambda
_{B}\mu ^{B}=-\mu _{B}\lambda ^{B}.  \label{32}
\end{equation}%
This means that the eigenvalues $\mathit{d}$ are imaginary, and the Dirac
operator is anti-Hermitian. The realness of the components of the bi-spinor $%
\widetilde{\eta }$ follows from the fact that all coefficients of the
secular equation Eq.(\ref{27}) after the Wick rotation $M\rightarrow iM$ are
real (if we cancel the imaginary unit on both sides). Since the Dirac
operator is anti-Hermitian, it follows that $\mathit{D}^{2}$ is Hermitian
and negative definite, and therefore the Witten operator Eq.(\ref{26}) is
Hermitian and positive definite. Completing the full Wick rotation

\begin{equation}
t\rightarrow it,\sqrt{\beta }\rightarrow \sqrt{\beta },M\rightarrow iM,%
\mathit{P}^{\prime }\rightarrow i\mathit{P}^{\prime },  \label{33}
\end{equation}%
where $\mathit{P}^{\prime }$ are all canonical momenta except $\sqrt{\beta }$%
, we can consider the functional integral Eq.(\ref{22}) to be well defined.

Thus, the congruence parameters of the spatial cross-sections in the
no-boundary wave function construction of the universe should be taken to be
$M(x)$. These cross-sections still converge to a single point as $%
t\rightarrow 0$, but now the boundary conditions are as follows:

\begin{equation}
M\rightarrow \infty ,\frac{d}{dt}\left( M_{kB}^{A}-\frac{1}{3}M\sigma
_{kB}^{A}\right) \rightarrow 0,\frac{d}{dt}\phi _{\alpha }\rightarrow 0.
\label{34}
\end{equation}%
As a result, the invariant propagator Eq.(\ref{22}) defines a no-boundary
wave function of the universe $\Psi \left[ M_{kB}^{A},\phi _{\alpha }\right]
$ in the privileged reference frame $\chi $. The wave function is the same
in each reference frame in accordance with the principle of covariance, but
now it means the validity of the principle of relativity.

Another issue is resolved by the Wick rotation Eq.(\ref{33}), which makes
the modified Witten operator Eq.(\ref{26}) positive definite. A Hermitian
positive-definite operator on a compact manifold $\Sigma $ has a discrete
spectrum. Thus, the class of privileged reference frames in a closed
universe is discrete.

\section{CONCLUSIONS}

Witten's identity underlies the positive-energy theorem for an island mass
distribution in Einstein's theory of gravity. In the case of a closed
universe, it allows one to reformat gravitational constraints into operator
equation Eq.(\ref{11}) on a spatial section $\Sigma $. Local gravitational
constraints serve as canonical generators of generally covariant
transformations, and the spectrum of the operator $W$ generates a discrete
class of privileged reference frames with a single wave function for the
universe. Thus, the principle of general covariance is transformed into the
principle of relativity for this class of reference frames. A representation
of the covariant wave function for the universe is proposed as a Euclidean
functional integral with a boundary condition different from the no-boundary
Hartle-Hawking wave function: instead of a zero radius for the universe at
the pole, an infinite asymptotic behavior of the Hubble parameter is
specified. This allows us to regularize the functional integral by Wick
rotation in the complex plane $M$. The proposed construction allows for a
generalization in which the closed universe is endowed with a nonzero proper
mass. This mass is not related to the energy distribution of physical
fields, but interacts gravitationally with this distribution. Dark matter
possesses these properties (its physical carriers have not yet been
discovered). However, this question requires a further research.

\section{ACKNOWLEDGEMENTS}

We are thanks V.A. Franke for useful discussions.

%%%\noindent $^{\ast }$ E-mail address: alex.lukyan@rambler.ru

%%%\noindent $^{+}$ E-mail address: inna.lukyan@mail.ru

%\begin{references}

\end{document}